\documentclass[10pt,prl,aps,twocolumn,showpacs,superscriptaddress]{revtex4}
\usepackage{graphicx}
\usepackage{amssymb}
\usepackage{floatflt}
\usepackage{bm}
\usepackage{color}

\newcommand{\bc}{\begin{center}}
\newcommand{\ec}{\end{center}}
\def\ba#1{\begin{array}{#1}\displaystyle}
\newcommand{\ea}{\end{array}}

\newcommand{\beq}{\begin{equation}}
\newcommand{\eeq}{\end{equation}}
\newcommand{\beqa}{\begin{eqnarray}}
\newcommand{\eeqa}{\end{eqnarray}}

\newcommand{\bi}{\begin{itemize}}
\newcommand{\ei}{\end{itemize}}

\def\lt#1{\left#1}
\def\rt#1{\right#1}
\def\t#1{\tilde{#1}}

\def\frc#1#2{\frac{#1}{#2}}

\newcommand{\bra}{\langle}
\newcommand{\ket}{\rangle}

\newcommand{\Tr}{{\rm Tr}}

\newcommand{\ep}{\epsilon}

\begin{document}

\title{Universal corrections to the entanglement entropy in gapped quantum spin chains: a numerical study}

\author{Emanuele Levi}
\affiliation
{Department of Mathematics, City University London, Northampton Square, EC1V 0HB}
\author{Olalla A. Castro-Alvaredo}
\affiliation
{Department of Mathematics, City University London, Northampton Square, EC1V 0HB}
\author{Benjamin Doyon}
\affiliation
{Department of Mathematics, King's College London, Strand WC2R 2LS }

\date{\today}


\begin{abstract}
We carry out a numerical study of the bi-partite entanglement entropy in the gapped regime of two paradigmatic quantum spin chain models: the Ising chain in an external magnetic field and the anti-ferromagnetic XXZ model. The universal scaling limit of these models is described by the massive Ising field theory and the $SU(2)$-Thirring (sine-Gordon) model, respectively. We may therefore exploit quantum field theoretical results to predict the behaviour of the entropy. We numerically confirm that, in the scaling limit, corrections to the saturation of the entropy at large region size are proportional to $K_0(2mr)$ where $m$ is a mass scale (the inverse correlation length) and $r$ the length of the region under consideration. The proportionality constant is simply related to the number of particle types in the universal spectrum. This was originally predicted in  \cite{entropy,entropy2} for two-dimensional quantum field theories. Away from the universal region our numerics suggest an entropic behaviour following quite closely the quantum field theory prediction, except for extra dependencies on the correlation length.

\end{abstract}

\maketitle

{\bf Introduction and discussion.} Entanglement is a fundamental property of the state of a quantum system. In the context of quantum computation, it is a crucial resource \cite{review}. Conceptually, it characterizes the structure of quantum fluctuations in a more universal way than other widely studied objects such as correlation functions. Developing theoretical measures of entanglement is therefore important to further understand the structure of quantum states, a problem of particular interest and difficulty for quantum many-body systems. A popular measure of entanglement is the bi-partite entanglement entropy. It measures the entanglement between two
complementary sets of observables in a quantum system \cite{bennet}. Interestingly, the entanglement entropy exhibits universal behaviour near quantum critical points: it has features which do not depend on the details of the model but rather on its universality class. In the last decade, this property has made the study of the entanglement entropy a very active field of research.

A fertile testing ground for these ideas
is the study of quantum spin chains. A quantum spin chain is a one-dimensional array of particles with spin degrees of freedom and, for our purposes, with nearest- (or few-nearest-) neighbour interactions (this is a local spin chain). Quantum spin chains have been realized in experiments \cite{qsc}. They provide ideal toy models for the study of a whole range of physical phenomena, as they describe interacting many-body systems yet are simple enough to allow for the computation of many quantities. This is especially true for integrable spin chains, since integrability gives rise to the complete characterization of the energy spectrum and states through techniques such as the coordinate and algebraic Bethe ansatz (see e.g.~\cite{Faddeev}). Quantum spin chains are extensively studied in
the context of quantum information science \cite{bose} and their bipartite entanglement for blocks of consecutive spins has
been investigated in many works (see e.g.~\cite{Latorre1,Latorre2,Latorre3,Jin,Lambert,KeatingM05,Weston}) both numerically and analytically. Most of these works concentrate on exact critical points.

 At the same time, in  1+1-dimensional quantum field theory (QFT) several universal results have been obtained for the entanglement entropy. Two of these results deserve special attention. First, the logarithmic growth of entanglement as the block size increases in conformal field theory (CFT),  which is controlled by the central charge \cite{Holzhey,CalabreseCardy}. This behaviour has been verified for many critical quantum chains \cite{Latorre1, Latorre2, Latorre3, Jin, Weston}. Second, the approach to saturation in massive QFT, which is controlled by the mass spectrum \cite{entropy,entropy2}. This has not been observed yet in quantum spin chains.

In the present work we investigate the entanglement entropy in two paradigmatic spin chain models (the Ising model and the XXZ model) {\em near to}, but not at, criticality. We numerically analyze for the first time the universal behaviour of the entanglement entropy of a block of length $L$ in the {\em near-critical scaling limit}. This scaling limit is expected to be described by {\em massive} QFT. In particular, we confirm massive QFT predictions \cite{entropy,entropy2}.

The entanglement entropy is the von Neumann entropy of the
reduced density matrix of a state $|\Psi\ket$ with
respect to a tensor factor of the Hilbert space ${\cal H}= {\cal A} \otimes {\cal B}$: \beq
\label{def}
    S = -\Tr_{\cal A}(\rho_A\,\log\rho_A)\quad \text{with} \quad \rho_A = \Tr_{\cal B}|\Psi\ket\bra\Psi|.
\eeq
Let $|\Psi\ket$ be the ground state of a quantum spin chain. In general, it is expected that, thanks to locality, the entanglement entropy of a continuous block $A$ of length $L$ saturates as $L\to\infty$: it only receives contributions from entanglement between spins surrounding boundary points of $A$. But this is not so at second order phase transitions. Assume that the Hamiltonian is parametrized by $h$,
and that there exists a value $h_c$ corresponding to a
quantum second order phase transition (critical point). These critical points are particularly interesting, as they characterise collective quantum behaviours.
At $h=h_c$, and in the thermodynamic limit where the length of the chain $N\rightarrow\infty$, the correlation length $\xi$ of the ground state is infinite, and the gap $\propto \xi^{-1}$ between the ground state and the continuum of excited states vanishes: the system is critical.
If the dynamical exponent is $z=1$ (so that the dispersion relation is linear), then the macroscopic, low-energy properties are universal and described by a CFT, see FIG.~1.

For a block $A$ of
length $L$, the entanglement entropy $S$ diverges for large
$L$ as \cite{Holzhey,CalabreseCardy}:
\beq
S(L)\sim \frc{c}3\log\left(\frac{L}{\epsilon}\right)+c_2,
\label{cft}
\eeq
where $c$ is the central charge of the CFT, $\epsilon$ is a short-distance scale (here and below taken as the inter-site spacing) and $c_2$ is a non-universal constant. The divergence occurs because with $\xi=\infty$ sites far apart are entangled.
Thus we may extract the value of the central charge by studying the divergence of the entanglement entropy in critical spin chains.

\begin{floatingfigure}{5cm}
  \includegraphics[width=4cm]{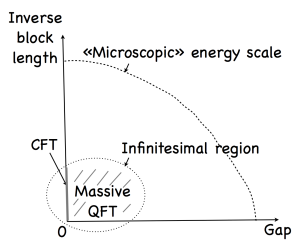}\\
  \caption{Quantum field theory reproduces the universal scaling limit of physical quantities: observation energies and the gap between the ground state and the first excited state are both very small as compared to the microscopic energy scales, like inter-site interaction energies.}
\end{floatingfigure}

In obtaining (\ref{cft}), the order of limits, first $h\to h_c$ then $L\gg \ep$, is very important. Instead,
let us take $h\to h_c$ and $L \gg \epsilon$ ``simultaneously": with $L/\xi=:mr$ fixed. This is the near-critical scaling limit, the shaded region depicted in FIG.~1. It is universal, described by a massive QFT corresponding to a perturbation off the original critical point. The mass scale $m$ may be identified with $\xi^{-1}$. In this limit, the gap of the spin chain is infinitesimally small as compared to microscopic energies (like the inter-site interaction energy), but the observation length is so large as to probe only the low-energy, universal, collective excitations of the quantum chain just above the gap; hence the gap still has an effect. The entanglement entropy then has the form of a ``saturation term'' $(c/3)\log(\xi/\ep)+c_1$ which diverges in the scaling limit, where $c_1$ is another non-universal constant, plus a universal scaling function $f(mr)$. The approach to saturation as $mr\to\infty$ is given by an exponential decay of $f(mr)$ which is solely determined by the spectrum of masses $\{m_i\}$ of the QFT \cite{entropy,entropy2}:
\beq
S=\frac{c}{3}\log\lt(\frc{\xi}\ep\rt) + c_1 -  \frc{1}8 \sum_i  K_0(2rm_i) + O\lt(e^{-3rm}\rt)\label{qft}
\eeq where $m\equiv m_1$ is the mass of the lightest particle. Here $K_0(z)$ is the modified Bessel
function. Thus, in the near-critical scaling limit, the entanglement entropy of the chain encapsulates information about the mass spectrum of the QFT. Further, despite the non-universality of $c_1$ and $c_2$, their difference {\em is} universal:
\begin{equation}\label{u}
    U:=c_1-c_2.
\end{equation}
In massive QFT, $U$ is related to the expectation value of a branch point twist field, as described in \cite{entropy}.
In contrast to the large number of works dealing with critical spin chains, the absence of studies of the near-critical scaling regime precludes a comparison with massive QFT results. The present work intends to fill this gap.

\section{} \vspace{-1.5cm}

{\bf Ising model.} This model has Hamiltonian
\begin{equation}\label{ising}
 H=-\frac{J}{2}\sum_{i=1}^N \left(\sigma^x_{i}\sigma^x_{i+1} + h \sigma_i^z\right),
\end{equation}
where $\sigma_i^a$ are Pauli matrices
acting on site $i$ and we consider periodic boundary conditions $\sigma_{i+N}\equiv \sigma_i$. The ``microscopic" energy scale indicated in FIG.~1 is the coupling constant $J>0$, and $h$ plays the role of an external magnetic field.
It is well-know (see e.g.~\cite{book}) that this model has a quantum critical point for $h=h_c=1$ described by a CFT with $c=\frac{1}{2}$.
The exact correlation length of the chain is given by $\xi^{-1}=\log h$ \cite{baxter} and, as expected, diverges at the critical point.
For $h>1$ the system is gapped and the ground state is unique (no symmetry is broken). Taking the scaling limit described above, a massive QFT is obtained: the free massive relativistic Majorana fermion.

A numerical study of the Ising model has the advantage that
a free fermion map can be used to perform computations in the thermodynamic regime.
This technique is used and explained in much detail in \cite{Latorre2} and it is based on the use of Toeplitz determinants \cite{Jin}.
We can then carry out extremely precise numerics on an exactly infinite chain for very large values of $L$ and very small values of $\xi^{-1}$, which are numerically inaccessible for more complex models.

A numerical computation at the conformal critical point yields the expected behaviour (\ref{cft}). For the Ising model, the exact value of $U=-0.131984...$
(we omit here the exact expression)
was first evaluated in \cite{entropy}. With $c_1={\log2}/{2}$ calculated in \cite{peschel} this gave, using (\ref{u}), the constant $c_2=0.478558...$. This was found to agree extremely well with the numerical results obtained in \cite{Latorre2}. Our numerics also confirm this value, see FIG.~2.
 \begin{figure}[h!]
  \includegraphics[width=8cm]{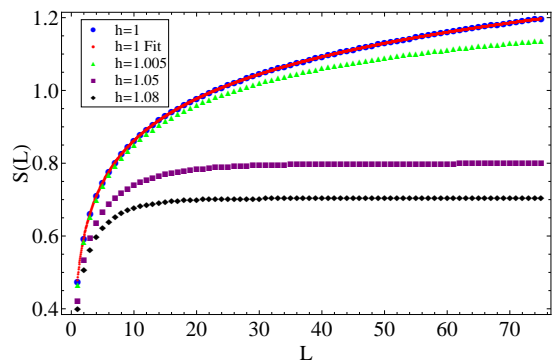}\\
  \caption{(color online) Entanglement entropy of the Ising model for several values of $h$.
  At the critical point $h=1$, the entropy is very well described by the function (\ref{cft}) with $c=0.500003$ and $c_2=0.478551$ (red dots). Away from
  the critical point we observe the rapid saturation of the entropy to a constant value.}
\end{figure}

Away from criticality, the spectrum of the QFT has a single particle. Hence formula (\ref{qft}), with only one term in the sum,
 should describe the large-$mr$ behavior in the limit $h\rightarrow 1^+$.
In our numerics, we take finite, increasing values of $\xi$, and each time the large-$L$ behaviour is fitted to
 \beq
S=\frac{1}{6}\log\xi + c_1(\xi)-  \frc{1}{\alpha(\xi)}  K_0(2L/\xi) \label{qft2}
\eeq
(we choose $\ep=1$ and use $mr=\frac{L}{\xi}$). We expect $c_1(\infty)=c_1$ and $\alpha(\infty)=8$.
FIG.~3 provides an example of such a fit with $\xi=40$, showing excellent agreement between numerical results and the Bessel function form of the correction to saturation.
\begin{figure}[h!]
  \includegraphics[width=7.5cm]{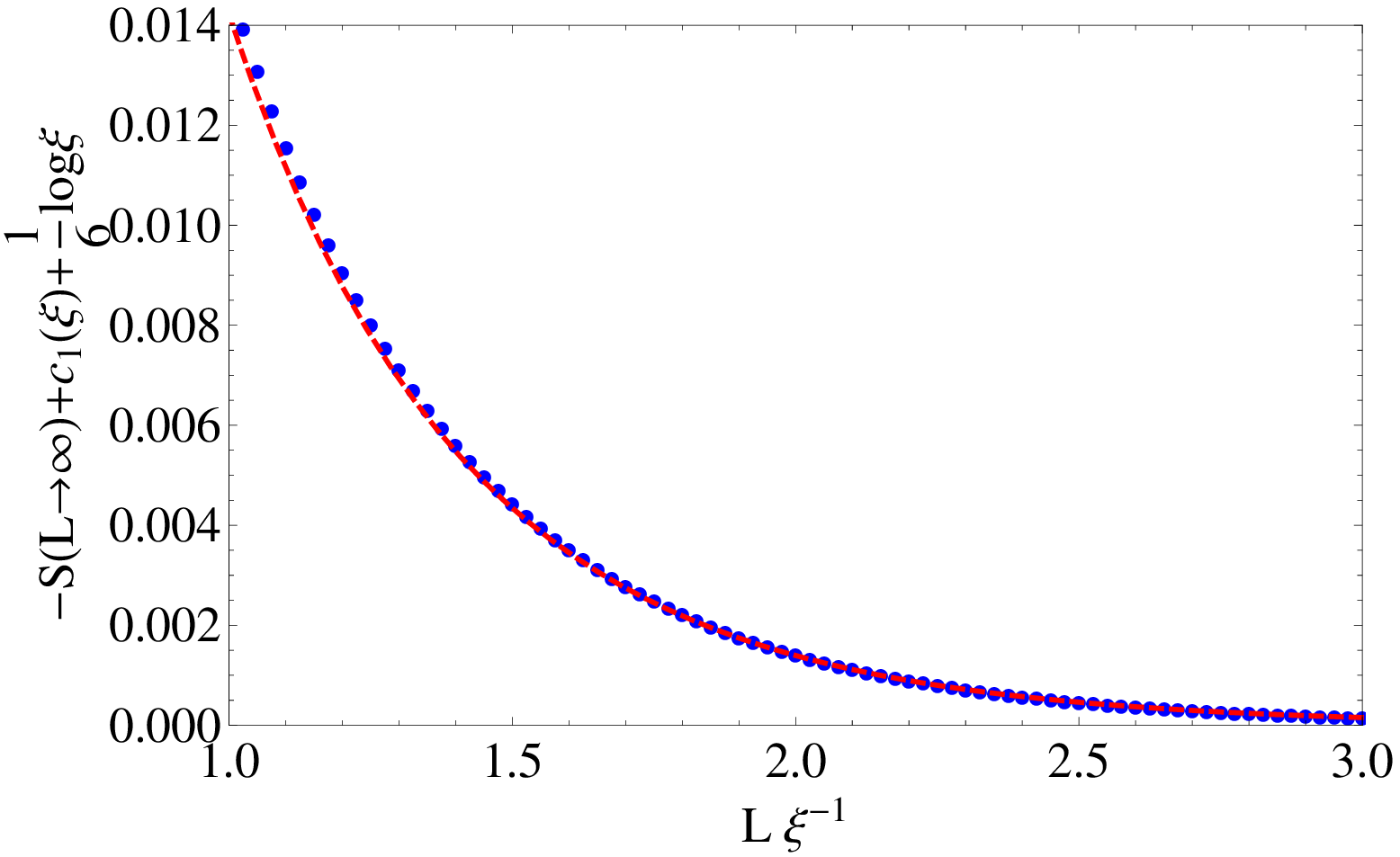}\\
  \caption{(color online) Correction to the saturation of the entanglement entropy of the gapped Ising chain for $\xi=40$. The solid circles represent the numerical values after subtraction of the exact saturation terms. The dashed red line represents the function $\frc18 K_0(2 L \xi^{-1})$.  The agreement is striking and clearly improves for increasing values of $L \xi^{-1}$.}
\end{figure}
FIG.~4 shows the data for $c_1(\xi)$.  The points agree very well with the exact function of $\xi$ predicted in \cite{peschel,CalabreseCardy}, lending support to our method.
\begin{figure}[h!]
  \includegraphics[width=7cm]{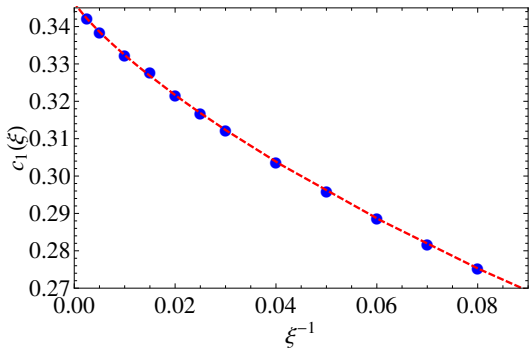}\\
  \caption{(color online) The circles are numerical values of $c_1(\xi)$ obtained by fitting (\ref{qft2}).
The dashed line represents exactly twice the function in equation (13) of \cite{peschel}, as expected since Peschel's work considers a partition into two semi-infinite regions. The agreement is again extremely good.}
\end{figure}
Finally, FIG.~5 illustrates the main result for the Ising chain. We show the function $\alpha(\xi)=:\alpha_x(\xi)$ obtained from a fit of (\ref{qft2}) in regions $L>x \xi$ for $x=1.5,2,2.5,3,3.5,4$ and 4.5. In all cases, $\alpha_x(\xi)$ appears constant as a function of $\xi$, and as $x$ increases this constant approaches $\alpha(\infty)=8$.
\begin{figure}[h!]
  \includegraphics[width=8cm]{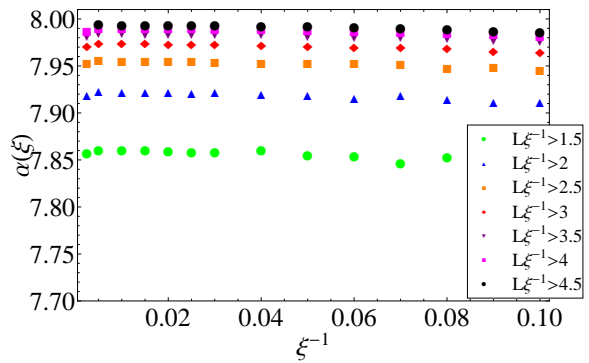}\\
  \caption{(color online) Numerical values of $\alpha_x(\xi)$. Linear fittings of the seven sets of data provide the asymptotic values $\alpha_{1.5}(\infty)=7.85581$, $\alpha_2(\infty)=7.9186$, $\alpha_{2.5}(\infty)=7.95252$, $\alpha_3(\infty)=7.97121$, $\alpha_{3.5}(\infty)=7.9818$, $\alpha_4(\infty)=7.98783$ and $\alpha_{4.5}(\infty)=7.99146$. Extrapolating in $x$ gives the prediction $\lim_{x\to\infty}\alpha_x(\infty) =7.99281$. }
\end{figure}\\

{\bf XXZ spin 1/2 model in the gapped regime.}
Let us now consider the Hamiltonian (with $J>0$)
\begin{equation}\label{Hxxz}
    H=J\sum_{i=1}^N\left(\sigma^x_{i}\sigma^x_{i+1}+\sigma^y_{i}\sigma^y_{i+1}+\Delta \sigma^z_{i}\sigma^z_{i+1}\right),
\end{equation}
with periodic boundary conditions. This is the
anti-ferromagnetic spin-$\frac{1}{2}$ XXZ chain (anisotropic Heisenberg model). This model displays a rich variety of features, depending on the value of the anisotropy parameter $\Delta$. In the thermodynamic limit $N\rightarrow \infty$ the model has a critical line on $\Delta\in[-1,1]$. For $\Delta\in(-1,1]$, including the Heisenberg model at $\Delta=1$, the thermodynamic limit is well described by a CFT (a free massless boson) with $c=1$. The other critical point $\Delta=-1$ is not described by CFT but rather corresponds to a critical theory with an infinitely degenerate ground state; the entanglement of such states has been studied in \cite{fractal}. For $\Delta>1$ the model is gapped, hence the scaling limit should be described by a massive QFT. The QFT is
the $SU(2)$-Thirring model (the sine-Gordon model at a special value of its coupling constant -- different values of the coupling are recovered by approaching the critical line in other ways, see e.g. \cite{otherways,ravanini}). This model has a spectrum of two asymptotic particles of equal mass.

In the XXZ model the exact correlation length is given by $ \xi(\Delta)^{-1}=\frac{\gamma}{2}+\sum_{n=1}^\infty \frac{(-1)^n}{n}\tanh(n\gamma)$,
where $\gamma=\cosh^{-1}(\Delta)$ \cite{baxter}.
\begin{figure}[h!]
  \includegraphics[width=8cm]{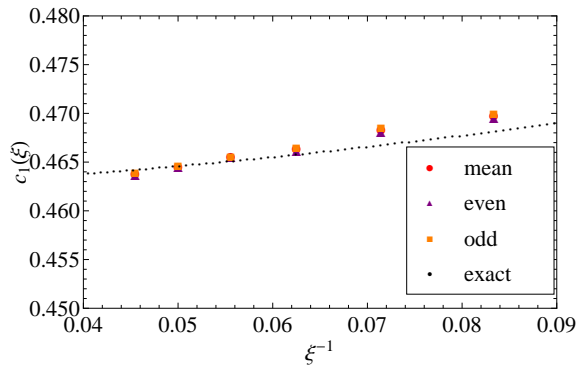}
  \caption{(color online) The function $c_1(\xi)$ near the critical point. The points represent the numerical results; the dotted line, the exact function as predicted in \cite{ravanini,ccp}. Denoting by $\t U(\xi)$ the function (14) of \cite{ravanini} minus the leading logarithmic term $\frac{1}{6}\log \xi$, the dotted line above is the function $c_1(\xi) = 2 \t U(\xi) + \log 2$. The factor 2 is due to the presence of two boundary points. The term $\log 2$ is related to the degeneracy: our zero-momentum ground state is composed in equal parts of two N\'eel-like orthogonal states of the same entropy. This function has the property $c_1(\infty)={2 \log2}/{3}$.}
\end{figure}

Numerical simulations on the XXZ chain were obtained by employing the density matrix renormalization group (DMRG) approach \cite{white}.
In contrast to the Ising example,  a new set of technical challenges arises. Whereas in the Ising model we could use formulae for the entropy where the length of the chain was infinity from the outset, in the XXZ case
we have to deal with finite chains.
This means that if we want to get meaningful results in the scaling limit we need to
ensure that $N\gg L$ and, at the same time, if we want to test the behaviour (\ref{qft}) we have to
consider $L\gg \xi$ whilst $\xi\gg \ep$ is sufficiently large so as to be close to the critical point $\Delta=1$.

We have found that for a given $\Delta$ an optimal choice is achieved by setting $N=5 \xi$. Longer chains are too hard to simulate whilst for shorter chains boundary effects make it impossible to obtain meaningful results. We considered the cases $\xi = 12, 14, 16, 18, 20$ and 22, and used again the form (\ref{qft2}). We fitted over the range $\xi<L<2\xi$.

A further challenge is posed by the fact that we have to consider periodic boundary conditions to recover the behaviour (\ref{qft2}). This makes the convergence of our DMRG algorithm much slower, forcing us to consider up to 1000 states to observe good convergence.

In order to guarantee the correct identification of the ground state we have employed two control parameters: the local magnetization $\langle\sigma_i^z\rangle$
and the truncation error. We used the condition $\langle\sigma_i^z\rangle=0$ as guiding principle for selecting the ground state. Imposing this condition is quite challenging for periodic chains. In our work we have been  able to guarantee $\langle\sigma_i^z\rangle<10^{-2}$.
\begin{figure}[h!]
  \includegraphics[width=7.5cm]{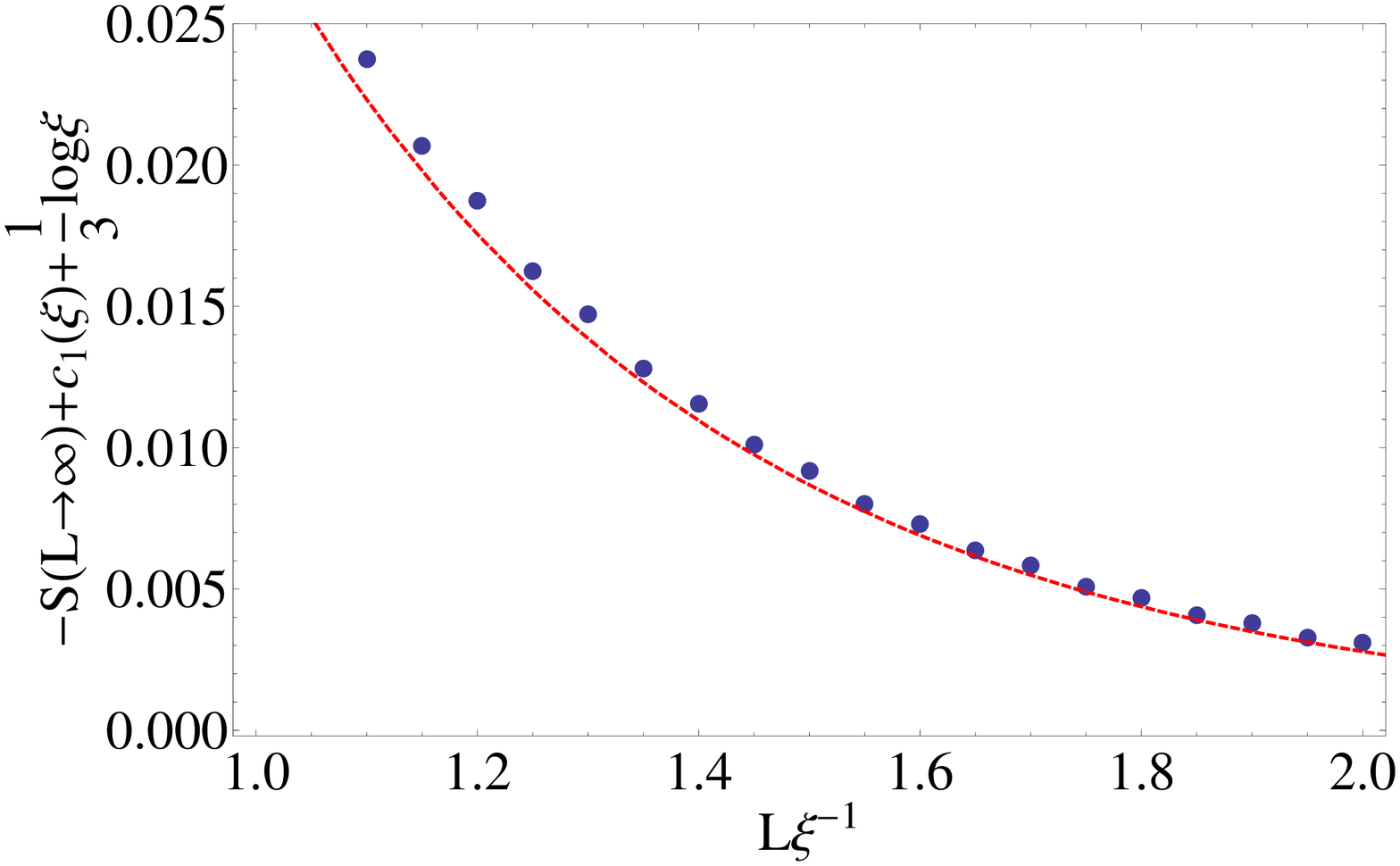}
  \caption{(color online) Correction to the saturation of the entanglement entropy of the gapped XXZ chain for $\xi=20$. The solid circles represent the numerical values after subtraction of the exact saturation terms. The dashed red line represents the function $\frc14 K_0(2 L \xi^{-1})$.
The agreement is good and clearly improves for increasing values of $L \xi^{-1}$.}
\end{figure}

For $L \gg \xi$ we observe small oscillations in the entropy between blocks with odd and even numbers of sites. These oscillations are slightly visible in the coefficient $c_1(\xi)$, FIG.~6, and the Bessel function, FIG.~7, and clearly visible in the function $\alpha(\xi)$, FIG.~8. Oscillations in the entropy have been seen for finite chains (see e.g.~\cite{essler}) where they arise as a consequence of finite size effects, albeit not for the von Neumann entropy. Clearly, in our case these oscillations must have a different origin. We believe that they are a consequence of the degeneracy of the ground state: for $\Delta>1$ the ground state is two-fold degenerate \cite{baxter,deg} in the thermodynamic limit (diagonalizing the momentum, two eigenvalues 0 and $\pi$ occur). Since we are always dealing with finite chains we should never see this degeneracy in our numerical simulations (the state with momentum eigenvalue 0 is the ``true'' ground state). However for long chains, the gap between the two states narrows and the precision of our algorithm fails to distinguish the energies of those states.
The oscillations should then be due to an alternate targeting of two linear combinations of the true (0-momentum) ground state and the $\pi$-momentum state (with small amplitude), depending on the parity of L.
\begin{figure}[h!]
  \includegraphics[width=8cm]{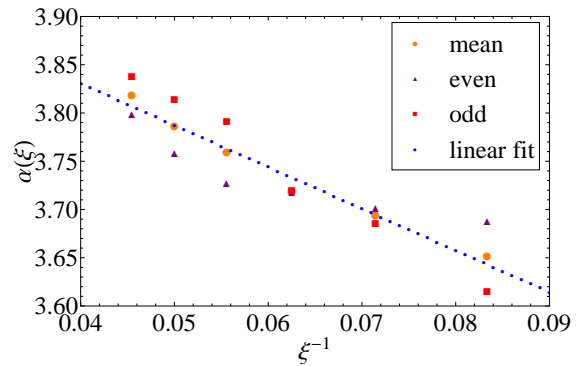}
  \caption{(color online) Values of $\alpha(\xi)$ approaching the critical point. The linear expression $\alpha(\xi)=-4.3261 \xi^{-1} + 4.0039$ reasonably fits the points and its extrapolation gives $\alpha(\infty)=4.0039$, which is quite close to the QFT prediction $\alpha^{\text{QFT}}=4$.}
\end{figure}

FIG.~8 shows that the coefficient $\alpha(\xi)$ converges with great precision to its theoretical value $4$.
Contrary to the Ising case, $\alpha(\xi)$ has a dependence on the correlation length. FIG.~7 shows a good fit to a Bessel function. Interestingly, this fit continues to be good away from $\Delta=1$ as long as the coefficient $\alpha(\xi)$ is changed as dictated by FIG.~8. This suggests that the function (\ref{qft2}) may provide to a good accuracy a universal description of the entropy of gapped spin chains.

Finally, at the critical point, we have found that the entropy is very well fitted by the function (\ref{cft}) with $c=1.00024$ and $c_2=0.733758$. We can therefore make a prediction for the universal QFT constant $U=c_1(\infty)-c_2=-0.27166$ in the $SU(2)$-Thirring model.

{\bf Conclusions.}  By studying gapped spin chains we have provided strong numerical evidence for the behaviour (\ref{qft}) suggested in \cite{entropy, entropy2} for two-dimensional QFT. Besides confirming QFT predictions, our results show that computing the entanglement entropy may be a good numerical way to determine the number of lightest particles in the universal spectrum, especially in combination with the knowledge that $\alpha(\infty)$ in (\ref{qft2}) must be a fraction of 8. Our results also suggest that some of the QFT predictions may still hold, with small changes, beyond the scaling regime.
It would be very interesting to derive this, or the appropriate modification of (\ref{qft2}), for spin chains from first principles.
We have made a prediction for the universal constant $U$ in the $SU(2)$-Thirring model, still to be confirmed by means of QFT techniques. Finally, it would be very interesting to carry out a similar analysis for non-integrable spin chains and thus verify the main result of \cite{entropy2}, namely, the fact that the behaviour (\ref{qft}) also holds for 1+1 dimensional non-integrable QFTs.

{\bf Acknowledgments.} In our study of the XXZ model we have used the numerical recipes provided by ALPS \cite{alps}. E.~Levi
wishes to thank M.~Troyer and A.~Feiguin from the ALPS team for their invaluable advice.

\end{document}